\documentclass[a4paper,fleqn]{cas-dc}

\usepackage[authoryear]{natbib}

\def\tsc#1{\csdef{#1}{\textsc{\lowercase{#1}}\xspace}}
\tsc{WGM}
\tsc{QE}

\begin{document}
\let\WriteBookmarks\relax
\def\floatpagepagefraction{1}
\def\textpagefraction{.001}

\shorttitle{DMC-Net for Pancreas Segmentation}    

\shortauthors{Jin Yang, Daniel S. Marcus and Aristeidis Sotiras}  

\title [mode = title]{DMC-Net: Lightweight Dynamic Multi-Scale and Multi-Resolution Convolution Network for Pancreas Segmentation in CT Images}  



%

\author[1]{Jin Yang}
\ead{yang.jin@wustl.edu}

\author[1]{Daniel S. Marcus}

\author[1,2]{Aristeidis Sotiras}


\ead{aristeidis.sotiras@wustl.edu}
\cortext[2]{Corresponding author}
\cormark[2]


\affiliation[1]{organization={Mallinckrodt Institute of Radiology},
            addressline={Washington University School of Medicine in St. Louis}, 
            city={St. Louis},
            postcode={63110}, 
            state={MO},
            country={USA}}

\affiliation[2]{organization={Institute for Informatics, Data Science and Biostatistics},
            addressline={Washington University School of Medicine in St. Louis}, 
            city={St. Louis},
            postcode={63110}, 
            state={MO},
            country={USA}}


\begin{abstract}
\hfill\break
\textbf{Background and Objective:} Convolutional neural networks (CNNs) have shown great effectiveness in medical image segmentation. However, they may be limited in modeling large inter-subject variations in organ shapes and sizes and exploiting global long-range contextual information. This is because CNNs typically employ convolutions with fixed-sized local receptive fields and lack the mechanisms to utilize global information. \\
\\
\textbf{Methods:} To address these limitations, we developed Dynamic Multi-Resolution Convolution (DMRC) and Dynamic Multi-Scale Convolution (DMSC) modules. Both modules enhance the representation capabilities of single convolutions to capture varying scaled features and global contextual information. This is achieved in the DMRC module by employing a convolutional filter on images with different resolutions and subsequently utilizing dynamic mechanisms to model global inter-dependencies between features. In contrast, the DMSC module extracts features at different scales by employing convolutions with different kernel sizes and utilizing dynamic mechanisms to extract global contextual information. The utilization of convolutions with different kernel sizes in the DMSC module may increase computational complexity. To lessen this burden, we propose to use a lightweight design for convolution layers with a large kernel size. Thus, DMSC and DMRC modules are designed as lightweight drop-in replacements for single convolutions, and they can be easily integrated into general CNN architectures for end-to-end training. The segmentation network was proposed by incorporating our DMSC and DMRC modules into a standard U-Net architecture, termed Dynamic Multi-scale and Multi-resolution Convolution network (DMC-Net). \\
\\
\textbf{Results:} To evaluate their effectiveness, we conducted experiments on pancreas segmentation from abdominal computed tomography (CT) images with the DMC-Net. The results demonstrate that our proposed DMSC and DMRC can enhance the representation capabilities of single convolutions and improve segmentation accuracy. Furthermore, their lightweight design led to lower computational complexity while maintaining or improving segmentation performance. \\
\\
\textbf{Conclusions:} The DMC-Net outperformed the state-of-the-art methods on pancreas segmentation in CT images.
\end{abstract}


\begin{highlights}
\item 2D and 3D Dynamic Multi-scale and Multi-resolution Convolution network for pancreas and pancreatic tumor segmentation
\item Dynamic Multi-Scale Convolution (DMSC) and Dynamic Multi-Resolution Convolution (DMRC) to dynamically capture features at varying scales and adaptively utilize global contextual information
\item The lightweight design for DMSC and DMRC to reduce computational complexity.
\item The proposed method achieves superior segmentation performance over state-of-the-art methods on pancreas segmentation in CT images.
\end{highlights}


\begin{keywords}
 \sep Dynamic Convolution \sep Multi-scale Convolution \sep Multi-resolution Convolution \sep Lightweight Network \sep Pancreas Segmentation
\end{keywords}

\maketitle
\section{Introduction}
Medical image segmentation is a critical task in clinical practice as it enables clinicians to locate target organs, quantify their anatomy, understand their morphology, and detect anatomical changes, thereby enabling accurate diagnosis and treatment planning. However, manual pixel-level labeling is time-consuming and error-prone. Therefore, automated segmentation tools are in high demand. Deep learning has revolutionized the field of image segmentation, and deep learning-based methods have been successful at troubleshooting semantic segmentation and instance segmentation problems \cite{chen2016semantic,noh2015learning,badrinarayanan2017segnet,hariharan2014simultaneous,isensee2021nnu}. In particular, convolutional neural networks (CNNs) have emerged as a robust, effective, and popular tool in automatic medical image segmentation, due to their high representation power, fast inference, and weight-sharing properties. \cite{unet,zhou2018unet++,cciccek20163d}. Indeed, CNNs have become the de facto standard for object segmentation in various medical imaging modalities, e.g., magnetic resonance (MR) \cite{milletari2016v}, computed tomography (CT) \cite{ibragimov2017segmentation,yang2023abdominal}, positron emission tomography (PET) \cite{zhao2018tumor}, and X-ray \cite{narin2021automatic}. \\
\\
The pancreas is a crucial organ for digestion and glucose metabolism, and pancreatic cancer is the fourth leading cause of cancer-related death due to its inferior prognosis \cite{mcguigan2018pancreatic}. Thus, accurate segmentation of pancreas and pancreatic masses is essential for the diagnosis and treatment of pancreas diseases. However, pancreas segmentation from abdominal CT images is challenging due to the substantial inter-patient variations in shape and size. Additionally, the pancreas only occupies a tiny part of the image. To solve this problem, several CNN-based methods have been proposed in recent years due to their superior segmentation capabilities \cite{kumar2019automated}. However, existing applications for pancreas segmentation rely heavily on multi-stage cascaded CNN frameworks \cite{HN,fixedpoint,Cai2017ImprovingDP,zhao2019fully,zhang2021automatic}. These frameworks typically employ CNNs (or other methods) in the first stage to make initial predictions and to identify a region of interest (ROI). In the second stage, CNNs are applied to make final dense predictions on that particular ROI, thus improving the overall segmentation accuracy of the single model. However, cascaded approaches lead to excessive and redundant use of computational resources and model parameters, as all models within the cascaded frameworks repeatedly extract similar low-level features. To enhance segmentation efficiency, one-stage CNN models have been proposed \cite{SCHLEMPER2019197,zhang2021deep}. However, these models employ convolutions with fixed-sized local receptive fields, thus limiting their ability to segment organs with large inter-patient variations in shape and size and to utilize global contextual information outside receptive fields. \\
\\
Several methodologies have been put forward to enhance convolutions with capabilities to capture features with varying shapes and scales. Feature pyramid methods were proposed to combine feature maps at different scales from different layers in a pyramidal hierarchic structure, enhancing the ability of the networks to make predictions on varying-scaled features \cite{lin2017feature,seferbekov2018feature,zhao2019m2det}. Additionally, multi-branch segmentation networks employed multiple segmentation paths to encode information at different scales \cite{wang2015object,aslani2019multi,chen2022mbanet}. However, these prior works focus on changing the macro-architecture of neural networks to use varying scaled features and may not be readily integrated into general CNN structures, which limits their applicability. Moreover, the implementation of these methods leads to a significant increase in the model size and computational complexity of the network, hindering their wide use for medical image segmentation. \\
\\
Other methods employ convolutional layers with both small and large kernel sizes to extract multi-scale features \cite{yang2023abdominal,he2019dynamic}. However, employing large convolutional kernels may lead to an increase in computational complexity. Previously, atrous convolutions with a large kernel size were applied since they can provide a large receptive field and alleviate computational burdens \cite{chen2017deeplab,hu2020automatic}. However, this is achieved by reducing the number of grid samples within receptive fields, thus reducing representation capabilities and lowering the accuracy of dense prediction. An alternative way is highly demanded to reduce computational complexity while maintaining or improving representation capabilities. \\
\\
Additionally, another limitation of CNNs lies in the inability of convolution layers to adaptively calibrate features using global contextual information. Some existing attention-based methods can tackle this problem \cite{hu2018squeeze,woo2018cbam,fu2019dual,zhong2020squeeze}. These methods were designed to model channel or spatial inter-dependencies among features, and then use the global information to selectively emphasize informative features and suppress less useful ones. However, these are designed as independent modules to improve the whole network rather than directly improving convolution layers, thus limiting their applicability and interpretability. \\
\\
In this paper, we aim to address these limitations by developing efficient solutions, specifically Dynamic Multi-Scale Convolution (DMSC) and Dynamic Multi-Resolution Convolution (DMRC). Both DMSC and DMRC modules are designed to enhance the capabilities of convolution layers to capture varying scaled features and utilize global contextual information to recalibrate local features. In the DMSC module, features at various scales are extracted by convolutions with different kernel sizes. Employing multiple convolutional layers increases the width of the network and enlarges the receptive field. Inspired by the recent works on large kernel-based CNNs \cite{ding2022scaling,guo2022segnext,lee20223d,yang2024d}, we propose to use depth-wise convolutions to reduce computational complexity. Subsequently, channel-wise dynamic mechanisms are used to model inter-dependencies between channel features extracted by two convolutions, thus calibrating these features via global information. In contrast, in the DMRC module, a convolution layer is employed on images with different resolutions to extract features at various scales, including pixel-wise and region-wise ones. Subsequently, the same dynamic mechanisms are applied to capture channel-wise global information. We designed our DMSC and DMRC as drop-in replacements of convolution layers, allowing for easy integration into existing CNN architectures without the need for changing their macro-architectures. CNNs equipped with DMSC or DMRC may be trained end-to-end for segmentation in the same manner as the original ones. \\ 
\\
To evaluate the effectiveness of our DMSC and DMRC in segmentation tasks, we integrated them into a general CNN architecture to generate our segmentation network, namely Dynamic Multi-scale and
Multi-resolution Convolution Network (DMC-Net). Specifically, a U-Net is used as the backbone \cite{unet}, and DMSC and DMRC are incorporated into it by replacing standard convolutions. Since U-Net employs standard convolution layers to extract fixed-scaled local features for segmentation without the utilization of global contextual information, it is straightforward to evaluate the benefits of our modules. To demonstrate these benefits, we evaluated our implementation on two commonly used abdominal CT pancreas segmentation benchmarks, namely the NIH The Cancer Imaging Archive Pancreas CT (TCIA-Pancreas) \cite{TCIA} for healthy pancreas segmentation and the Medical Segmentation Decathlon Pancreas challenge (MSD-Pancreas) \cite{antonelli2022medical} for the segmentation of the pancreas and pancreatic masses. Our results indicate that our segmentation network achieved superior segmentation performance compared to state-of-the-art methods. Importantly, DMC-Net using DMSC and DMRC modules achieved significant improvements over the standard U-Net baseline that uses standard convolutional layers. Lastly, the lightweight design of the convolutional layers led to lower computational complexity without compromising segmentation performance.\\
\\
Our contributions can be summarized as follows.

\begin{itemize}
    \item We developed two novel modules, Dynamic Multi-Scale Convolution (DMSC) and Dynamic Multi-Resolution Convolution (DMRC), to enhance the capabilities of convolutional layers. These modules dynamically capture features at varying scales and adaptively utilize global contextual information to recalibrate features. Both of them are designed as 2D and 3D drop-in blocks, allowing for easy integration into existing CNN architectures by replacing convolutional layers without the need for changing their macro-architectures. CNNs equipped with DMSC or DMRC can be trained end-to-end for segmentation.
    \item We proposed lightweight designs for the DMRC and DMSC modules. These designs can reduce the computational complexity of segmentation models while maintaining or improving their representation capabilities.
    \item We generated a 2D and 3D Dynamic Multi-scale and Multi-resolution Convolution network (2D-DMC-Net and 3D-DMC-Net) for automatic medical image segmentation by incorporating DMSC and DMRC modules into a standard 2D and 3D U-Net, respectively.
    \item Our experimental results in pancreas segmentation demonstrate that replacing convolutional layers with DMSC or DMRC in the U-Net architecture improves segmentation accuracy. Additionally, the proposed DMC-Net showed superior performance over state-of-the-art pancreas segmentation methods.
\end{itemize}

\section{Methods}
We propose 2D and 3D Dynamic Multi-Scale Convolution (DMSC) and Dynamic Multi-Resolution Convolution (DMRC) modules, and the incorporation of them into 2D and 3D backbones generate 2D and 3D Dynamic Multi-scale and Multi-resolution Convolution networks (2D-DMC-Net and 3D-DMC-Net). To simplify the presentation, we will focus on describing the 2D modules and the 2D network in this section and highlighting any potential differences between the 2D and 3D implementations at the end of the section.

\subsection{Network architecture}
\begin{figure*}[!t]
\centering
\includegraphics[width=\textwidth]{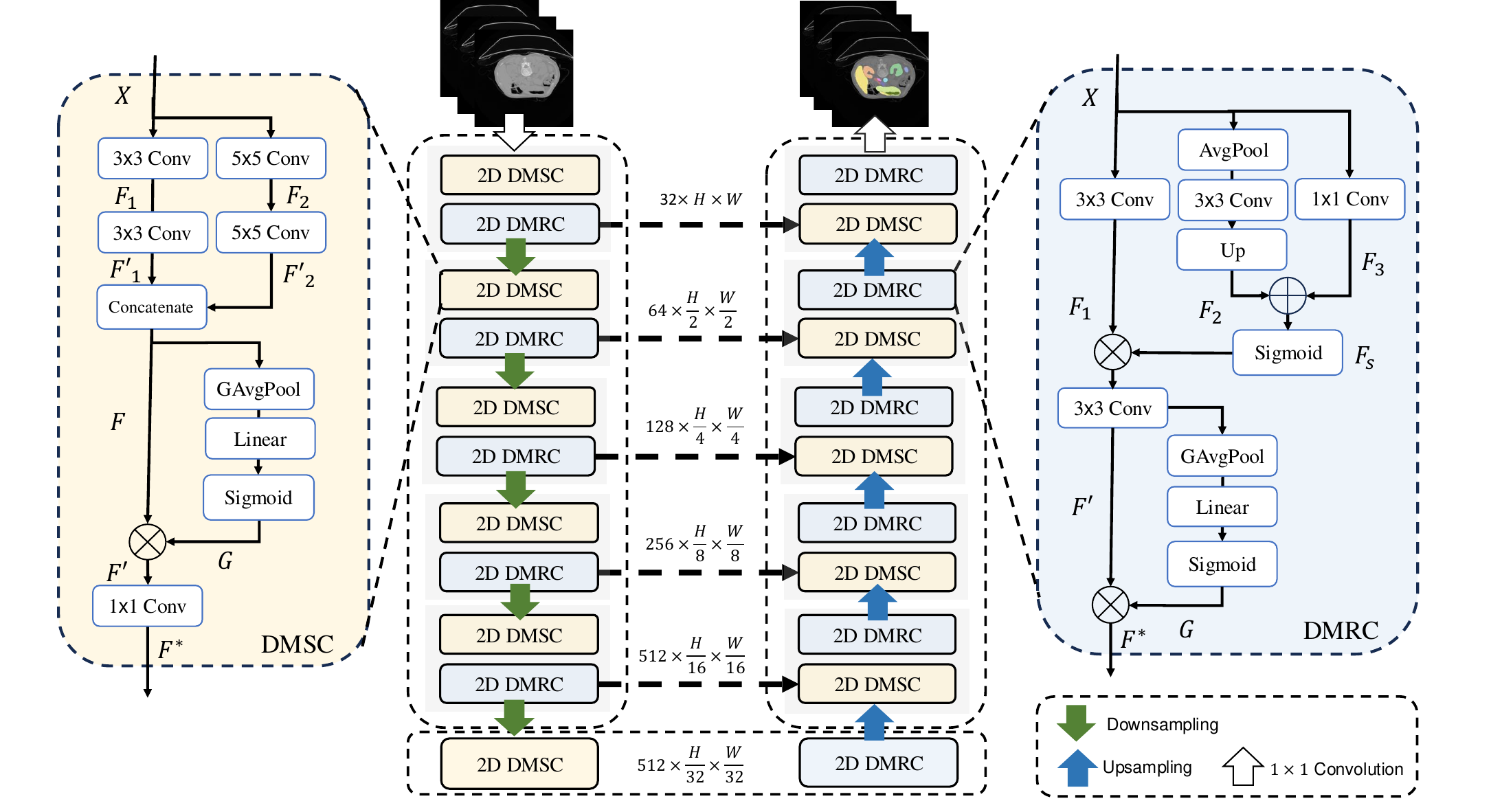}
\caption{The architecture of 2D DMC-Net. It is designed as a U-shaped network, consisting of a 5-stage encoder, a 5-stage decoder, and a bottleneck. In each stage, the basic block consists of a 2D DMSC module and a 2D DMRC module. A $1\times1$ convolutional layer is utilized to project input features to $32$ channels in the encoder, and another $1\times1$ convolutional layer is utilized to make pixel-wise predictions in the decoder.}
\label{fig:1}
\end{figure*}

We propose a Dynamic Multi-Scale and Multi-Resolution Convolution network (DMC-Net) for the automatic pancreas and pancreatic mass segmentation. This model was designed based on integrating DMSC and DMRC modules into a standard U-Net backbone \cite{unet}. Specifically, DMRC was placed after DMSC, and we used them to replace convolutional layers at each layer. $2\times2$ Max-pooling layers and $3\times3$ transposed convolutional layers were used to downscale and upscale the feature maps. They were also used to increase and decrease the number of feature maps, respectively. At the final layer, a $1\times 1$ convolution was used to generate the segmentation prediction. The proposed 2D-DMC-Net has 6 layers, and the number of feature maps in each layer was 32, 64, 128, 256, 512, and 512, respectively (Figure~\ref{fig:1}). The proposed 3D-DMC-Net has 5 layers with 32, 64, 128, 256, and 512 feature maps, respectively (Figure~\ref{fig:2}).

\subsection{Dynamic Multi-Resolution Convolution (DMRC)}
\begin{figure*}[!t]
\centering
\includegraphics[width=0.9\textwidth]{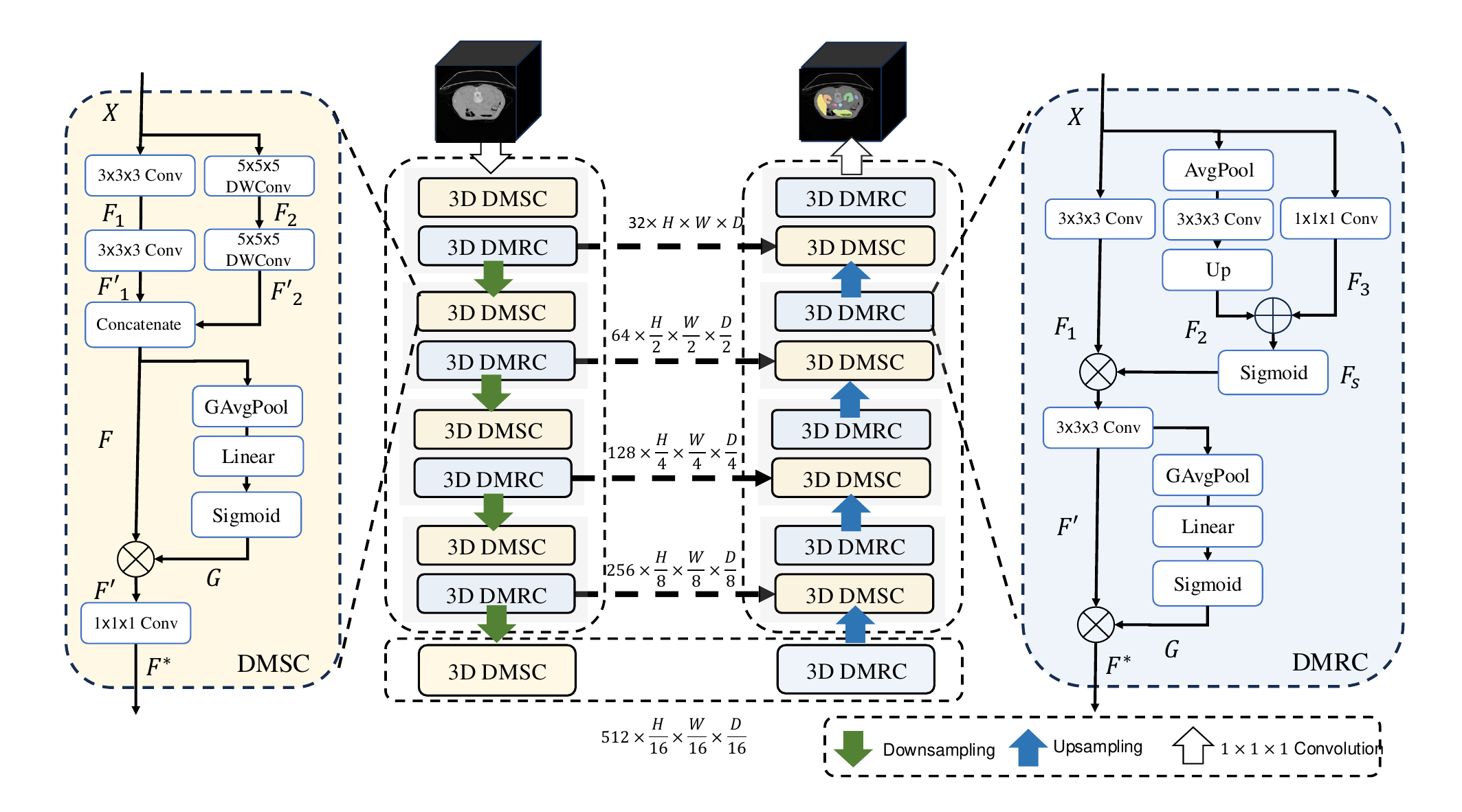}
\caption{The architecture of 3D DMC-Net. It is designed as a U-shaped network, consisting of a 4-stage encoder, a 4-stage decoder, and a bottleneck. In each stage, the basic block consists of a 3D DMSC module and a 3D DMRC module. A $1\times1\times1$ convolutional layer is utilized to project input features to $32$ channels in the encoder, and another $1\times1\times1$ convolutional layer is utilized to make voxel-wise predictions in the decoder.}
\label{fig:2}
\end{figure*}

We propose the Dynamic Multi-Resolution Convolution (DMRC) module to dynamically capture features at various scales. This is achieved by employing convolutions on images with different resolutions and adaptively utilizing global contextual information (2D: Figure~\ref{fig:1}; 3D: Figure~\ref{fig:2}). Specifically, three parallel paths are adopted to generate multi-scale features from input feature maps $\boldsymbol{X}=[\boldsymbol{x}_1,\boldsymbol{x}_2,...,\boldsymbol{x}_{C}]\in \mathbb{R}^{C\times H \times W}$. First, a convolutional layer with a kernel size of 3, denoted as $\mathcal{F}_{conv}(\cdot)$, is applied to extract features $\boldsymbol{F}_1\in \mathbb{R}^{C\times H \times W}$ from the input $\boldsymbol{X}$ with the original resolution:
\begin{equation}
    \nonumber
    \boldsymbol{F}_1 = \mathcal{F}_{conv}(\boldsymbol{X}).
    \label{Eq:1}
\end{equation}
The second path employs a convolutional layer to extract features at another scale $\boldsymbol{F}_2\in \mathbb{R}^{C\times H \times W}$ by utilizing neighboring contextual information from the input $\boldsymbol{X}$. Specifically, this path first utilizes average pooling $\mathcal{F}_{AvgPool}(\cdot)$ operation with $4\times4$ filter and $4\times4$ stride to capture and aggregate neighboring contextual information from input feature maps $\boldsymbol{X}$, effectively reducing their resolution by 4. Subsequently, a $3\times3$ convolution, denoted as $\mathcal{F}_{conv}(\cdot)$, is applied to generate features $\boldsymbol{F}_2\in \mathbb{R}^{C\times \frac{H}{4} \times \frac{W}{4}}$ from rescaled images. Thus, these features $\boldsymbol{F}_2$ are at different scales from features $\boldsymbol{F}_1$. Lastly, nearest neighbor interpolation $\mathcal{F}_{up}(\cdot)$ is used to upsample these features $\boldsymbol{F}_2$ to the original input dimension. These operations are summarized in the equation below:
\begin{equation}
    \nonumber
    \boldsymbol{F}_2 = \mathcal{F}_{up}(\mathcal{F}_{conv}(\mathcal{F}_{AvgPool}(\boldsymbol{X}))).
    \label{Eq:2}
\end{equation}
However, aggregating neighboring contextual information by average pooling may lead to the loss of pixel-wise information. This limitation is solved by the third path. This path employs a $1\times1$ convolution layer, denoted as $\mathcal{F}_{conv}^{(1)}(\cdot)$, to extract pixel-wise contextual information $\boldsymbol{F}_3\in \mathbb{R}^{C\times H \times W}$:
\begin{equation}
    \nonumber
    \boldsymbol{F}_3 = \mathcal{F}_{conv}^{(1)}(\boldsymbol{X}).
    \label{Eq:3}
\end{equation}
Subsequently, the pixel-wise information $\boldsymbol{F}_3$ and neighboring region-wise features $\boldsymbol{F}_2$ are fused to generate the features $\boldsymbol{F}_s\in \mathbb{R}^{C\times H \times W}$:
\begin{equation}
    \nonumber
    \boldsymbol{F}_s=\boldsymbol{F}_2 \oplus \boldsymbol{F}_3.
    \label{Eq:4}
\end{equation}
When generating multi-scale features $\boldsymbol{F}\in \mathbb{R}^{C\times H \times W}$ by fusing features in three paths $\boldsymbol{F}_1$ and $\boldsymbol{F}_s$, we utilize a spatial-wise dynamic mechanism. Specifically, a Sigmoid function $\mathcal{F}_{Sig}(\cdot)$ is used to convert features $\boldsymbol{F}_s$ to attention values, which are then utilized to calibrate features $\boldsymbol{F}_1$ via element-wise multiplication:
\begin{equation}
    \nonumber
    \boldsymbol{F}=\boldsymbol{F}_1 \otimes\mathcal{F}_{Sig}(\boldsymbol{F}_s).    
    \label{Eq:5}
\end{equation}
In the subsequent step, another $3\times3$ convolution layer $\mathcal{F}^{(3)}_{conv}(\cdot)$ is utilized to extract organ-specific spatial features and generate feature maps $\boldsymbol{F}'\in \mathbb{R}^{C\times H \times W}$:
\begin{equation}
    \nonumber
    \boldsymbol{F}' = \mathcal{F}^{(3)}_{conv}(\boldsymbol{F}).
    \label{Eq:6}
\end{equation}
We utilize dynamic mechanisms to capture global information $\boldsymbol{G}\in \mathbb{R}^{C\times 1 \times 1}$ by modeling channel-wise inter-dependencies among features $\boldsymbol{F}'$. This is achieved by cascading a global average pooling $\mathcal{F}_{AvgPool}(\cdot)$, a Linear layer $\mathcal{F}_{Conv}^{(1)}(\cdot)$, and a Sigmoid activation $\mathcal{F}_{Sig}(\cdot)$: 
\begin{align}
    \nonumber
    \boldsymbol{G} = \mathcal{F}_{Sig}(\mathcal{F}_{Conv}^{(1)}(\mathcal{F}_{GAvgPool}(\boldsymbol{F}'))).  
\end{align}
Finally, the output features of the DMRC module $\boldsymbol{F}^*\in \mathbb{R}^{C_{out}\times H \times W}$ are generated by using this global information $\boldsymbol{G}$ to dynamically calibrate multi-resolution features $\boldsymbol{F}'$: 
\begin{align}
    \nonumber
    \boldsymbol{F}^*=\boldsymbol{F}' \otimes \boldsymbol{G}. 
\end{align}
3D implementation details: There are a few differences between 2D and 3D implementations that we would like to highlight. Let $\boldsymbol{X}=[\boldsymbol{x}_1,\boldsymbol{x}_2,...,\boldsymbol{x}_{C}]\in \mathbb{R}^{C\times D\times H \times W}$ be input feature maps. In the second path, average pooling operations with $1\times4\times4$ filter and $1\times4\times4$ stride are applied for $\boldsymbol{X}$.

\subsection{Dynamic Multi-Scale Convolution (DMSC)}
We propose the Dynamic Multi-Scale Convolution (DMSC) module to dynamically capture varying-scaled features. This is achieved by employing convolutions with multi-scale kernels and adaptively utilizing global contextual information (2D: Figure~\ref{fig:1}; 3D: Figure~\ref{fig:2}). Two parallel paths are adopted to capture features at varying scales from input feature maps $\boldsymbol{X}=[\boldsymbol{x}_1,\boldsymbol{x}_2,...,\boldsymbol{x}_{C}]\in \mathbb{R}^{C\times H \times W}$. Convolutions with different kernel sizes of 3 and 5, denoted as $\mathcal{F}_{conv}^{(3)}(\cdot)$ and $\mathcal{F}_{conv}^{(5)}(\cdot)$, are employed in each path (see Eq.~\ref{Eq:9} and Eq.~\ref{Eq:10}). Application of convolutions in parallel paths results in their increased width to capture finer-grained features, thus granting DMSC enhanced representation capabilities over the single convolution. Moreover, our design employs receptive fields of various sizes, which helps convolutions extract features at different scales.
\begin{align}
    \boldsymbol{F}_1 = \mathcal{F}_{conv}^{(3)}(\boldsymbol{X}).
    \label{Eq:9}
\end{align}
\begin{align}
    \boldsymbol{F}_2 = \mathcal{F}_{conv}^{(5)}(\boldsymbol{X}). 
    \label{Eq:10}
\end{align}
Subsequently, two additional convolutions, $\mathcal{F}_{conv}^{(3)}(\cdot)$ and $\mathcal{F}_{conv}^{(5)}(\cdot)$, are employed to generate feature maps $\boldsymbol{F}'_1\in \mathbb{R}^{C\times H \times W}$ and $\boldsymbol{F}'_2\in \mathbb{R}^{C\times H \times W}$ (see Eq.~\ref{Eq:11} and Eq.~\ref{Eq:12}). This design increases the depth of the DMSC module, providing a larger receptive field and enhancing its capability to capture richer features.
\begin{align}
    \boldsymbol{F}'_1 = \mathcal{F}_{conv}^{(3)}(\boldsymbol{F}_1).
    \label{Eq:11}
\end{align}
\begin{align}
    \boldsymbol{F}'_2 = \mathcal{F}_{conv}^{(5)}(\boldsymbol{F}_2). 
    \label{Eq:12}
\end{align}
Afterwards, feature maps $\boldsymbol{F}'_1$ and $\boldsymbol{F}'_2$ are concatenated along the channel dimension to form features $\boldsymbol{F}\in \mathbb{R}^{2C\times H \times W}$:
\begin{align}
    \nonumber
    \boldsymbol{F} = [\boldsymbol{F}'_1;\boldsymbol{F}'_2] 
\end{align}
To enhance the capabilities of DMSC to utilize global contextual information, dynamic mechanisms are employed. Specifically, global information $\boldsymbol{G}\in \mathbb{R}^{2C\times 1 \times 1}$ is extracted from features $\boldsymbol{F}$ by cascading a global average pooling $\mathcal{F}_{AvgPool}(\cdot)$, a Linear layer $\mathcal{F}_{Conv}^{(1)}(\cdot)$, and a Sigmoid activation $\mathcal{F}_{Sig}(\cdot)$: 
\begin{align}
\nonumber
    \boldsymbol{G} = \mathcal{F}_{Sig}(\mathcal{F}_{Conv}^{(1)}(\mathcal{F}_{GAvgPool}(\boldsymbol{F}))).  
\end{align}
This information can describe the importance of features $\boldsymbol{F}$, allowing the calibration of these local features by highlighting the more important ones and suppressing the less informative ones. The calibrated features $\boldsymbol{F}'\in \mathbb{R}^{2C\times H \times W}$:
\begin{align}
\nonumber
    \boldsymbol{F}'=\boldsymbol{F} \otimes \boldsymbol{G}.
\end{align}
To ensure the output features have the same dimension as the input ones, a channel reduction mechanism is required to reduce the number of channels to the original count $C$. Since local features have been calibrated by global contextual information, a $1\times1$ convolutional layer, denoted as $\mathcal{F}_{Conv}^{(1)}(\cdot)$, is utilized to select feature maps based on their importance. This channel information will guide the convolutional layer to preserve the important features while dropping less informative ones, thus generating output features $\boldsymbol{F}^*\in \mathbb{R}^{C\times H \times W}$:
\begin{align}
\nonumber
    \boldsymbol{F}^* = \mathcal{F}_{Conv}^{(1)}(\boldsymbol{F}'). 
\end{align}

\subsection{Lightweight design}
To extract multi-scale features in the DMSC module, convolutional layers with a kernel size of $5\times5$ are employed. This implementation leads to an increase in computational complexity. To lessen this burden, we propose using depth-wise convolution designs (DWConv) \cite{liu2022convnet} to construct the DMSC module as a lightweight solution (Figure \ref{fig:2}). We apply this lightweight design to 3D DMSC modules by replacing $5\times5\times5$ standard convolutional layers with $5\times5\times5$ depth-wise convolutional layers $\mathcal{F}_{DWConv}^{(5)}$ to extract features $\boldsymbol{F}_2$ and $\boldsymbol{F}'_2$ from input features $\boldsymbol{X}$:
\begin{align}
\nonumber
    \boldsymbol{F}_2 = \mathcal{F}_{DWConv}^{(5)}(\boldsymbol{X}).
\end{align}
\begin{align}
\nonumber
    \boldsymbol{F}'_2 = \mathcal{F}_{DWConv}^{(5)}(\boldsymbol{F}_2).
\end{align}
Convolutional layers or depth-wise convolutional layers with larger kernel sizes provide a larger receptive field than those with a kernel size of 5. However, they may result in a lower segmentation performance and significantly increase computational complexity. \\
\\
Lastly, to enable convolutional layers to extract features from input images with different sizes, we applied average pooling operations with $4\times4$ filters and $4\times4$ stride in 2D DMRC modules and with $1\times4\times4$ filters and $1\times4\times4$ stride in 3D DMRC modules. This is an optimal lightweight design because it can improve segmentation performance while decreasing computational complexity. In contrast, using average pooling operations with $2\times2$ filters and $2\times2$ stride in 2D DMRC modules, and with $1\times2\times2$ filters and $1\times2\times2$ in 3D DMRC modules leads to higher computational complexity and potentially lower segmentation accuracy.

\section{Experiments and Results}
\subsection{Datasets}
We conducted experiments on two publicly available datasets. The first dataset used was the National Institutes of Health (NIH) repository of normal pancreas CTs, which is made publicly available through The Cancer Imaging Archive (TCIA-Pancreas) \cite{TCIA}. This dataset comprises 80 contrast-enhanced 3D abdominal CT scans obtained from patients who had neither major abdominal pathologies nor pancreatic cancer lesions. The original CT scans have a resolution of $512 \times 512$ pixels with varying pixel sizes and slice thickness between 1.5 $mm$ and 2.5 $mm$. The number of slices ranges from $181$ to $466$ varying from patient to patient. Manual annotations of the pancreas were conducted slice-by-slice by a medical student and were then verified and refined by an experienced radiologist. These annotations were used as the segmentation ground truth for this dataset. \\
\\
Additionally, we utilized the public abdominal CT pancreas dataset from the Medical Segmentation Decathlon challenge (MSD-Pancreas; MICCAI 2018) \cite{antonelli2022medical}, which was acquired at the Memorial Sloan Kettering Cancer Center (New York, NY, USA). This dataset consists of 420 portal-venous phase CT scans of patients undergoing resection of pancreatic masses and is divided into 281 training and 139 testing scans. The raw training CT scans have resolutions of $512 \times 512$ pixels with varying pixel sizes and slice thickness. The number of slices in the axial view ranges from 37 to 751. Manual annotations were performed for both pancreatic parenchyma and pancreatic mass (i.e., cyst or tumor), and three semantic classes were provided, namely pancreas, pancreatic mass, and background. In our experiments, we used only the training samples since these were the only ones that were accompanied by annotations.

\subsection{Implementation details}
The DMC-Net was implemented using PyTorch. We used a combination of cross-entropy loss $\mathcal{L}_{CE}$ and dice loss $\mathcal{L}_{Dice}$. The loss function $\mathcal{L}$ can be formulated between the prediction $\hat{y}$ and the target $y$ as:
\begin{align}
    \nonumber
    \mathcal{L}=\mathcal{L}_{CE}(\hat{y},y)+\mathcal{L}_{Dice}(\hat{y},y).
\end{align}
During training, the stochastic gradient descent (SGD) optimizer was used for model optimization. The initial learning rate was set to 0.001 with a Nesterov momentum of 0.99 and it was decayed with a poly learning rate scheduler during training. The number of training epochs was set to 1000 for all training sessions. The batch size was 12 and 2 for 2D and 3D data, respectively. Models were trained and validated on NVIDIA Tesla A100 PCI-E Passive Single GPU each with 40GB of GDDR5 memory. The code will be made available from the lab Github upon acceptance for publication.\\
\\
Several pre-processing techniques were implemented for images from both two datasets. Firstly, volumetric scans were resampled to a resolution of $0.85\times 0.85\times1.0$ $mm^3$ and $0.8\times 0.8\times2.5$ $mm^3$ in TCIA-Pancreas and MSD-Pancreas datasets, respectively. Subsequently, we normalized intensity values by performing min-max normalization over the whole volume. Finally, data were prepared as 2D patches with size $512\times512$ $mm^2$ and 3D patches with size $192\times192\times96$ $mm^3$ in the TCIA-Pancreas dataset. Additionally, data were prepared as 2D patches with size $512\times512$ $mm^2$ and 3D patches with size $192\times192\times64$ $mm^3$ in the MSD-Pancreas dataset.\\
\\
Several data augmentation techniques were implemented. Patches were rotated along three axes between $[-30, 30]$ and scaled between $(0.7, 1.4)$ with a probability of 0.2. All patches were mirrored with a probability of 0.5 along all axes. Zero-centered additive Gaussian noise with the variance drawn from $U(0,0.1)$ was added to each voxel in the sample with a probability of 0.15. Gaussian blur was applied with a probability of 0.2 for each sample by a Gaussian kernel whose width is sampled from $U(0.5, 1.5)$. Brightness and contrast were applied to voxel intensities both with a probability of 0.15.\\
\\
We employed a 5-fold cross-validation approach to get stable and reliable evaluation results on both two benchmarks. Segmentation performance was evaluated by two metrics: Dice Similarity Coefficient (DSC) and $95\%$ Haudroff distance (95HD). The architecture complexity was evaluated by two metrics: the number of parameters (Params) and the number of Floating Point Operations (FLOPs). The Wilcoxon signed-rank test was implemented to statistically quantify the differences in segmentation performance.

\begin{table*}[h!]
\centering
\caption{Performance comparison between our approach and the state-of-the-art methods for pancreas segmentation on the NIH TCIA-Pancreas dataset. Segmentation performance was evaluated by DSC ($\%$).  The DSC is reported as mean±std. The best performance for each class of models is highlighted by \textbf{bold}.}
\resizebox{0.7\textwidth}{!}{
\begin{tabular}{c| c | c | c|c} 
 \hline
 Models & Methods & Mean DSC  & Max DSC  & Min DSC \\ 
 \hline
 2D & C-RNN& 82.4±6.7 &  90.1 & 60.0 \\ 

 & RSTN& 84.50±4.97 &  91.02 & 62.81 \\ 
 & TVMS-Net& 85.19±4.73 &  $\boldsymbol{91.42}$ & 68.69 \\ 
 & DMC-Net (ours) & $\boldsymbol{85.64}$±3.93 & 90.86 & $\boldsymbol{77.20}$ \\
  \hline
 2.5D & DeepOrgan& 71.8±10.7 &  86.9 & 25.0 \\ 

 & HNN & 81.14±7.30 &  89.98 & 44.69 \\ 

 & Fixed-point & 82.37±5.68 & 90.85 & 62.43 \\
 & DSD-ASPP-Net& 85.49±4.77 &  91.64 & 67.19 \\
 & DCNN & 84.90 &  91.02 & 62.81 \\
 & MADC-Net  & 86.49±1.44 & - & - \\
 & RTUNet & 86.25±4.52 &  $\boldsymbol{93.02}$ & 72.55 \\
  \hline
 3D & Attention U-Net & 81.48±6.23 &  - & - \\
 & CNN & 84.47±4.36 &  91.54 & 70.61 \\
 & ECTN & 85.58±3.98 &  91.30 & 72.88 \\ 
 & DMC-Net (ours) & $\boldsymbol{87.97}$±3.52 & 91.56 & $\boldsymbol{77.64}$ \\ 
 \hline
\end{tabular}}
\label{table:1}
\end{table*}

\begin{table*}[h!]
\centering
\caption{Performance comparison between our approach and state-of-the-art methods for pancreas and pancreatic tumor segmentation on the MSD-Pancreas dataset. Segmentation performance was evaluated by Mean DSC ($\%$). Our experiments were implemented using 5-fold cross-validation following the same partition as nnU-Net \cite{isensee2021nnu}. The best and second performances are highlighted by \textbf{bold} and \underline{underline}.}
\resizebox{0.7\textwidth}{!}{
\begin{tabular}{c | c | c | c | c } 
 \hline
 Models & Methods & Pancreas DSC & Tumor DSC & Average DSC \\
 \hline
 2D & nnU-Net& 77.38 & 35.01 & 56.19 \\
 & DMC-Net (ours) & 79.82 & 42.07 & 60.95 \\ 
 \hline
 3D & V-Net & 79.01 &  35.99 & 57.50 \\ 
 & V-NAS & 79.9 &  38.78 & 59.34 \\
 & C2FNAS & 80.59  & 52.87 & 66.73 \\
 & nnU-Net & \underline{82.14} & 54.28 & \underline{68.21} \\
 & SAR & 73.42  & 25.19 & 49.31 \\
 & Distill DSM  & 79.20  & 37.65 & 58.43 \\
 & DiNTS & 81.02  & \underline{55.35} & 68.19 \\
 & HyperSegNAS  & 79.98  & 54.88 & 67.43 \\
 & DMC-Net (ours) & $\boldsymbol{82.92}$ & $\boldsymbol{60.71}$ & $\boldsymbol{71.81}$  \\ 
 \hline
\end{tabular}}
\label{table:2}
\end{table*}

\subsection{Comparison to state-of-the-art frameworks}
We compared the proposed model with state-of-the-art (SOTA) models on two datasets. These SOTA methods have been previously applied to segment the pancreas on the TCIA dataset and pancreatic tumors on the MSD dataset. \\
\\
On the NIH TCIA-Pancreas dataset, we compared 2D-DMC-Net with three 2D state-of-the-art models, namely C-RNN \cite{Cai2017ImprovingDP}, RSTN \cite{yu2018recurrent}, TVMS-Net \cite{chen2022pancreas}. Additionally, we compared 3D-DMC-Net with seven 2.5D state-of-the-art models, namely DeepOrgan \cite{deeporgan}, HNN \cite{roth2018spatial}, Fixed-points \cite{fixedpoint}, DSD-ASPP-Net \cite{hu2020automatic}, DCNN \cite{zhang2021automatic}, MADC-Net \cite{li20222}, and RTUNet \cite{qiu2023rtunet}, and three 3D state-of-the-art models, namely Attention U-Net \cite{SCHLEMPER2019197}, CNN \cite{zhang2021deep}, and ECTN \cite{zheng2023extension}. Table \ref{table:1} presents the comparison results of these methods. When our 2D-DMC-Net is compared with other 2D models, it outperforms these models and achieves the highest Mean DSC value, showing the effectiveness of our proposed model. Additionally, 2D-DMC-Net shows the lowest standard deviation in Mean DSC and the smallest gap between Max DSC and Min DSC, which demonstrates the robustness of our model in pancreas segmentation. Both 2.5D models and 3D models account for the 3D nature of volume data, so we compare our 3D DMC-Net with other 2.5D and 3D models. Our 3D-DMC-Net demonstrates a superior segmentation performance than other methods. Specifically, it shows the highest Mean DSC in pancreas segmentation. Moreover, it achieves the highest robustness than other methods due to the smallest standard deviation in Mean DSC and the smallest gap between Max DSC and Min DSC. \\
\\
On the MSD dataset, we only compared our 2D-DMC-Net with 2D nnU-Net \cite{isensee2021nnu}, since few 2D segmentation methods were applied to this dataset. We also compared our 3D-DMC-Net with eight 3D methods, namely, V-Net \cite{milletari2016v}, V-NAS \cite{zhu2019v}, C2FNAS \cite{yu2020c2fnas}, 3D nnU-Net \cite{isensee2021nnu}, SAR \cite{zhang2021sar}, Distill DSM \cite{maheshwari2021distill}, DiNTS \cite{he2021dints}, and HyperSegNAS \cite{peng2022hypersegnas}. Table \ref{table:2} presents the comparison results of these methods. Our 2D DMC-Net outperforms 2D nnU-Net on both pancreas and pancreatic tumor segmentation, thus achieving a superior performance in average segmentation from both two classes. Furthermore, when compared with other 3D methods, our 3D-DMC-Net demonstrated superior performance in both pancreas and pancreatic tumor segmentation. Specifically, it achieves the highest DSC in pancreas segmentation, pancreas tumor segmentation, and on average.

\begin{table*}[h!]
\centering
\caption{The results of the ablation study on TCIA-Pancreas and MSD-Pancreas datasets. Segmentation performance was evaluated using DSC ($\%$), and 95HD (mm). Evaluation results were reported as Mean±Std or Median [$25\%$ Q1,$75\%$ Q3]. All experiments were performed using 5-fold cross-validation with the same partition. The best and the second performances are highlighted by \textbf{bold} and \underline{underline}. ($^*$: $p<0.01$ with Wilcoxon signed-rank test between DMC-Net and each method.)}
\resizebox{1.\textwidth}{!}{
\begin{tabular}{>{\centering\arraybackslash}p{1cm}|>{\centering\arraybackslash}p{1.7cm}|>
{\centering\arraybackslash}p{1.5cm}|>
{\centering\arraybackslash}p{1.5cm}|>
{\centering\arraybackslash}p{1.5cm}|>
{\centering\arraybackslash}p{1.5cm}|>
{\centering\arraybackslash}p{1.5cm}|>
{\centering\arraybackslash}p{1.7cm}|>
{\centering\arraybackslash}p{1cm}}
 \hline
   \multirow{3}{*}{Models} & \multirow{3}{*}{Methods} & \multicolumn{2}{c|}{TCIA} & \multicolumn{5}{c}{MSD} \\
  \cline{3-9} & & Pancreas Mean DSC  & Pancreas Mean 95HD & Pancreas Mean DSC & Pancreas Median 95HD & Tumor Mean DSC & Tumor Median 95HD & Average Mean DSC \\
 \hline
 \multirow{4}{*}{2D} & UNet & 82.80 ±7.05 & 6.14 ±3.64 & 77.39 ±9.46 & 5.80 [4.61,9.24] & 36.72 ±30.84 & 14.98 [6.22,50.05] & 57.06 \\ 
  & DMRC-Net & 84.08 ±5.78 & 4.47 ±3.62 &  \underline{78.95} ±9.24 & \underline{4.86} [3.56,6.34] & \underline{39.81} ±30.60 & \underline{10.66} [5.49,45.56] & \underline{59.38} \\ 
  & DMSC-Net & \underline{84.39} ±6.30 & \underline{4.43} ±3.21 &  78.60 ±9.03 & 5.00 [4.12,6.52] & 38.28 ±30.39 & 13.71 [5.37,39.66] & 58.44  \\ 
  & DMC-Net & $\boldsymbol{85.64}^*$ ±3.93  & $\boldsymbol{3.80}^*$ ±1.28 &  $\boldsymbol{79.82}^*$ ±8.32 & $\boldsymbol{4.63}^*$ [3.42,6.50] & $\boldsymbol{42.07}^*$ ±25.20 & $\boldsymbol{10.06}^*$ [4.44,28.27] & $\boldsymbol{60.95}^*$  \\
 \hline
 \multirow{4}{*}{3D} & UNet & 83.56 ±6.43 & 5.06 ±4.74 & 80.34 ±8.37 & 4.63 [3.27,6.28] & 52.31 ±32.48 & 6.86 [3.38,19.88] & 66.32 \\ 
  & DMRC-Net & \underline{85.62} ±4.62 & 3.69 ±2.70 &  \underline{81.44} ±7.57 & 4.28 [3.29,5.32] & 57.10 ±31.09 & \underline{5.00} [3.2,12.83] & 69.27  \\ 
  & DMSC-Net & 85.67 ±4.86 & \underline{3.74} ±2.92 &  81.36 ±7.47 & \underline{4.20} [3.11,5.43] & \underline{57.37} ±30.03 & 5.25 [3.24,11.33] & \underline{69.36} \\ 
  & DMC-Net & $\boldsymbol{87.97}^*$ ±3.52 & $\boldsymbol{2.83}^*$ ±0.73 & $\boldsymbol{82.92}^*$ ±6.91 & $\boldsymbol{4.08}^*$ [3.04,5.13] & $\boldsymbol{60.71}^*$ ±27.02 & $\boldsymbol{4.50}^*$ [2.86,10.71] & $\boldsymbol{71.81}^*$ \\
 \hline
\end{tabular}}
\label{table:3}
\end{table*}

\subsection{Ablation study on DMRC and DMSC modules}
\begin{figure*}[!t]
\centering
\includegraphics[width=1.0\textwidth]{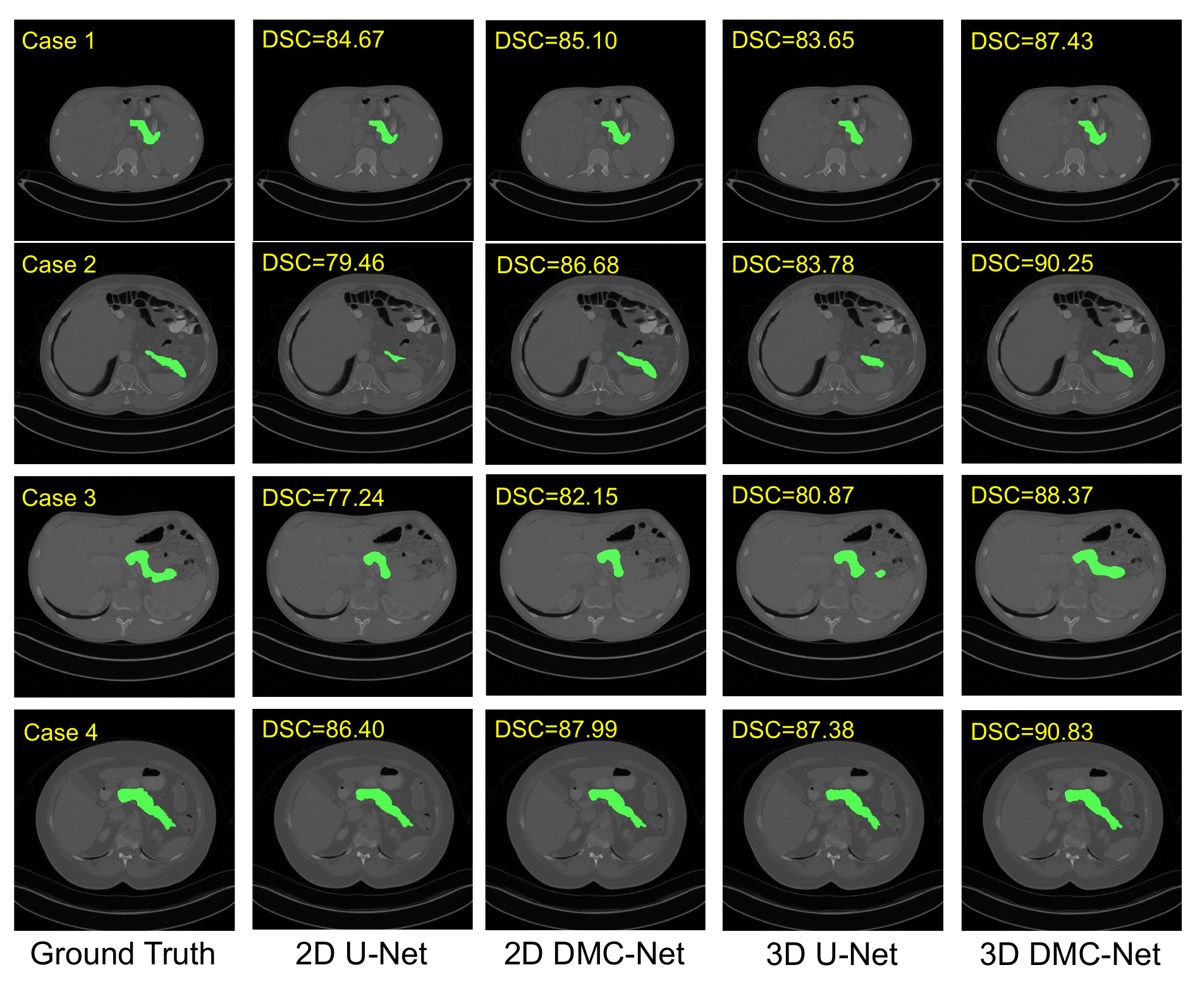}
\caption{Visualizations of the pancreas segmentation results on the NIH TCIA-Pancreas dataset. The pancreas is marked in green. 2D DMC-Net shows better segmentation quality than 2D U-Net, and 3D DMC-Net shows better segmentation quality than 3D U-Net.}
\label{fig:4}
\end{figure*}

\begin{figure*}[!t]
\centering
\includegraphics[width=1.0\textwidth]{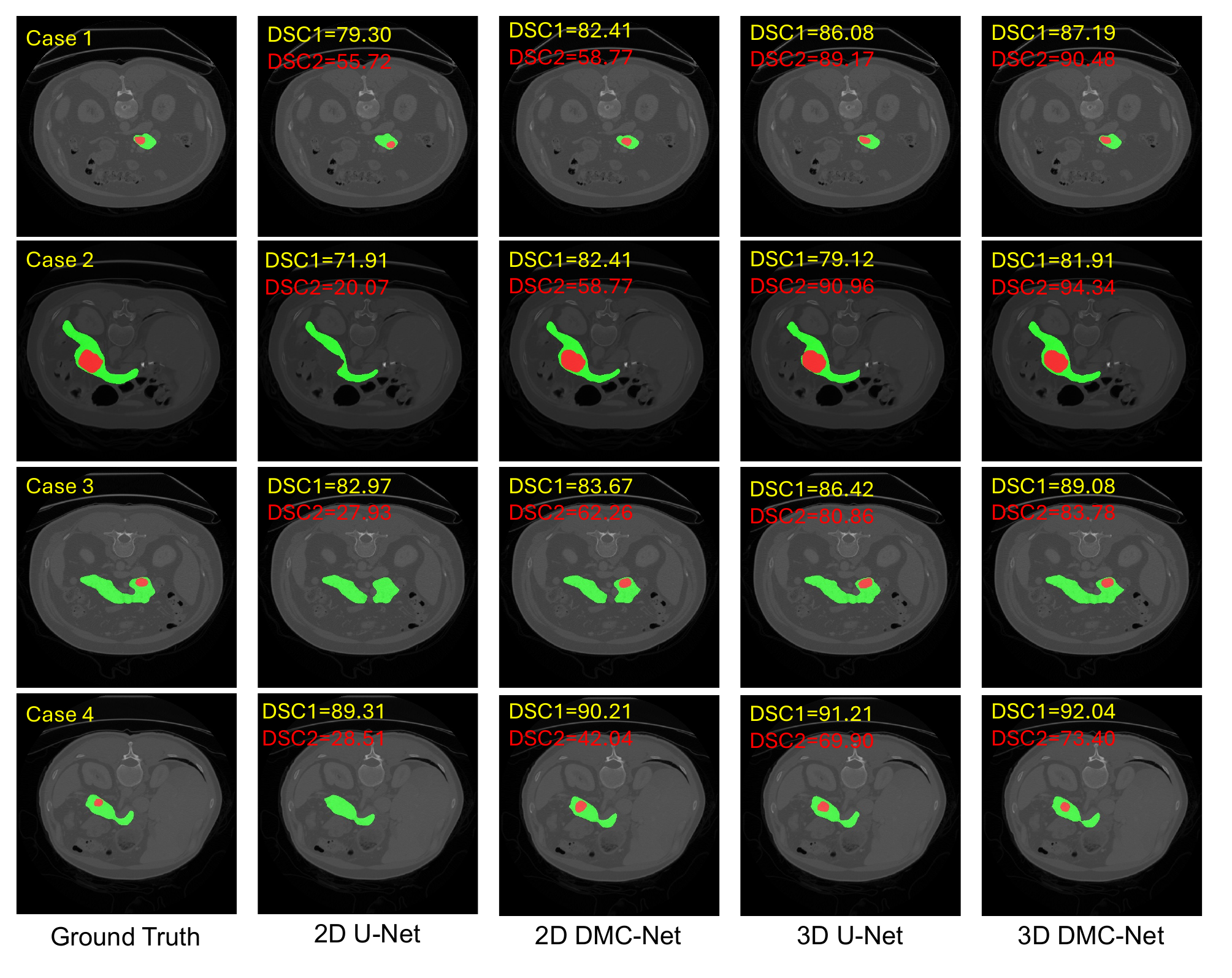}
\caption{Visualizations of the pancreas and pancreatic mass segmentation results on the MSD-Pancreas dataset. The pancreas is marked in green, and the tumor is marked in red. 2D DMC-Net shows better segmentation quality than 2D U-Net, and 3D DMC-Net shows better segmentation quality than 3D U-Net.}
\label{fig:5}
\end{figure*}

We conducted an ablation study on both 2D and 3D  DMC-Net to evaluate the contributions of our DMRC and DMSC modules on improving pancreas segmentation accuracy. This ablation study was performed using both NIH TCIA-Pancreas and MSD Pancreas datasets.\\
\\
We compared the segmentation performance between our DMC-Net and U-Net baselines. The U-Net was implemented following a similar design as DMC-Net in the number of layers and the number of feature maps in each layer. In this U-Net, each layer in both the encoder and decoder utilized a convolutional block consisting of two convolutional layers. For 2D implementations, each convolutional layer had a $3\times3$ kernel, while for 3D implementations, each convolutional layer had a $3\times3\times3$ kernel. The 2D DMRC-Net was implemented using our 2D DMRC module to replace the first convolutional layer in each convolution block in all layers. The 2D DMSC-Net was built using our 2D DMSC module to replace the first convolutional layer in each convolution block in all layers. The same design was implemented to 3D DMSC-Net and 3D DMRC-Net.\\
\\
Table \ref{table:3} presents the results of our ablation study. Specifically, we observed that incorporating either the DMRC or DMSC modules into both 2D and 3D U-Net (DMRC-Net and DMSC-Net) resulted in superior segmentation performance than the U-Net, as measured by DSC and 95HD on the TCIA-Pancreas dataset. Additionally, the DMC-Net, incorporating both modules, consistently outperformed all other networks, demonstrating that the integration of our DMRC and DMSC modules may enhance the segmentation performance of U-Net. Subsequently, we conducted a similar ablation study on the MSD-Pancreas dataset. Integrating either the DMRC or DMSC modules into both 2D and 3D U-Net (DMRC-Net and DMSC-Net) enhanced the segmentation capabilities of U-Net for both pancreas and pancreatic tumors. Implementing DMC-Net by integrating modules achieved the highest segmentation accuracy in terms of DSC and 95HD. Figure \ref{fig:4} and Figure \ref{fig:5} further illustrate the effectiveness of our modules in improving pancreas segmentation on the TCIA-Pancreas and MSD-Pancreas datasets, respectively.

\begin{table*}[h!]
\centering
\caption{The results of lightweight design analysis on the TCIA-Pancreas dataset. The segmentation performance was evaluated using DSC ($\%$) and 95HD (mm). The computational complexity of the networks was evaluated using Params (M) and FLOPs (G). All experiments were implemented using 5-fold cross-validation with the same partition. The best segmentation performance and lower computational complexity in each method are highlighted by \textbf{bold}.}
\resizebox{1.0\textwidth}{!}{
\begin{tabular}{>{\centering\arraybackslash}p{1.2cm}|>{\centering\arraybackslash}p{1.8cm}|>
{\centering\arraybackslash}p{3.4cm}|>
{\centering\arraybackslash}p{1.2cm}>
{\centering\arraybackslash}p{1.2cm}|>
{\centering\arraybackslash}p{1.2cm}>
{\centering\arraybackslash}p{1.5cm}} 
 \hline
 Models & Methods & Lightweight design & DSC & 95HD & Params & FLOPs \\ 
 \hline
 \multirow{6}{*}{2D} & U-Net & - & 82.80 & 6.14 & 22.80 & 94.86 \\
 \cline{2-7} & \multirow{2}{*}{DMRC-Net} & $2\times2$ AvgPool &    83.32 & 4.52 & 43.77 & 145.47\\
  & & $4\times4$ AvgPool (ours) &  $\boldsymbol{84.08}$ & $\boldsymbol{4.47}$ & 43.77 & $\boldsymbol{138.86}$ \\
  \cline{2-7} & \multirow{4}{*}{DMSC-Net} & $5\times5$ Conv (ours) & $\boldsymbol{84.39}$ & $\boldsymbol{4.43}$ & 89.61 & 343.13 \\
 & & $5\times5$ DWConv &  83.32 & 4.74 & $\boldsymbol{38.49}$ & $\boldsymbol{147.80}$ \\
 & &$7\times7$ Conv    &  83.26 & 4.72 &  139.89 & 536.81 \\
 & & $7\times7$ DWConv &  83.18 & 4.81 &  38.61  & 150.15 \\
 \hline
 \multirow{6}{*}{3D} & U-Net & - & 83.56 & 5.06 & 25.89 & 1398.81 \\
 \cline{2-7} 
 & \multirow{2}{*}{DMRC-Net} & $2\times2$ AvgPool &  $\boldsymbol{85.63}$ & 3.71 & 47.86 & 2098.93 \\
 & & $4\times4$ AvgPool (ours) & 85.62 & $\boldsymbol{3.69}$ & 47.86 & $\boldsymbol{2011.70}$ \\
 \cline{2-7}
 & \multirow{4}{*}{DMSC-Net} & $5\times5\times5$ Conv & 85.53 & 3.89 &  138.35 & 6766.37 \\
 & & $5\times5\times5$ DWConv (ours) & $\boldsymbol{85.67}$ & $\boldsymbol{3.74}$ & $\boldsymbol{41.02}$  & $\boldsymbol{2140.72}$ \\
 & & $7\times7\times7$ Conv   & 85.55 & 3.76 & 309.35 & 15061.43 \\
 & & $7\times7\times7$ DWConv & 85.54 & 3.78 & 41.66  & 2337.84 \\
 \hline
\end{tabular}}
\label{table:4}
\end{table*}

\subsection{Ablation study on lightweight designs}
Next, we conducted ablation studies to evaluate the importance of the lightweight design of DMRC and DMSC modules. First, we focused on assessing the importance of implementing the average pooling in 2D DMRC-Net using filters of size 4 and stride 4 ($4\times4$ AvgPool) by comparing it with an implementation that uses filters of size 2 and stride 2 ($2\times2$ AvgPool). Specifically, we evaluated the two architectures in terms of segmentation performance and computational complexity. Subsequently, we assessed the impact of the lightweight depthwise convolution in the 2D DMSC modules by comparing the segmentation performance and computational complexity of different DMSC-Nets. Specifically, we tested DMSC-Nets with $5\times5$ convolutional layers ($5\times5$ Conv), $5\times5$ depth-wise convolutional layers ($5\times5$ DWConv), $7\times7$ convolution layers ($7\times7$ Conv), and  $7\times7$ depth-wise convolution layers ($7\times7$ DWConv). The same ablation study was implemented for 3D DMRC and DMSC modules.\\
\\
Table \ref{table:4} presents the results of lightweight design analysis. The design of the 2D DMRC module showed that the DMRC-Net with $4\times4$ average pooling layers achieved better segmentation performance and lower computational complexity (higher DSC and 95HD values, and lower FLOPs). Similarly, using $4\times4\times4$ average pooling layers in the 3D DMRC module resulted in superior segmentation performance and lower computational complexity. Thus, this lightweight design was applied to both 2D and 3D DMRC modules.\\
\\
We did not apply the lightweight design to our 2D DMSC module. Specifically, evaluating this lightweight design of DMSC modules showed that 2D DMSC-Net equipped with $5\times5$ convolutional layers achieved the best segmentation performance. On the contrary, 3D DMSC-Net with the lightweight design ($5\times5\times5$ DWConv) demonstrated the highest segmentation performance and the lowest computational complexity. Thus, we applied the lightweight design to 3D DMSC modules and used $5\times5\times5$ depth-wise convolutional layers instead of $5\times5\times5$ convolutional layers.

\section{Conclusions}
We proposed a Dynamic Multi-scale and Multi-resolution Convolution network (DMC-Net) for the automatic pancreas and pancreatic mass segmentation from CT images. This network employs novel Dynamic Multi-Scale Convolution (DMSC) and Dynamic Multi-Resolution Convolution (DMRC) modules, whose goal is to improve the representation capabilities of standard convolutional layers by capturing features and various scales and utilizing global contextual information. Our experimental analysis demonstrated that the proposed modules may improve the segmentation performance of the network. An important component of the proposed architecture was the emphasis on adopting lightweight designs for the proposed convolutional layers. This facilitates the integration of the proposed modules into existing CNN architectures without requiring any changes in their macro-architectures. The experimental results showcased the superior performance of our 2D and 3D DMC-Net on two pancreas segmentation benchmarks compared to state-of-the-art methods. Lastly, we demonstrated that the use of the proposed DMSC and DMRC modules may improve the segmentation accuracy of CNNs without increasing the computational complexity compared to baseline CNN architectures.

\section*{Declaration of competing interests}
The authors declare that they have no known competing financial interests or personal relationships that could have appeared to influence the work reported in this paper.

\section*{Acknowledgments}
Computations were performed using the facilities of the Washington University Research Computing and Informatics Facility (RCIF). The RCIF has received funding from NIH S10 program grants: 1S10OD025200-01A1 and 1S10OD030477-01.

\section*{Funding}
This work was supported by NIH grant U24 CA258483.

\section*{CRediT authorship contribution statement}
JY: conceptualization, methodology, formal analysis, writing the original draft, reviewing, and editing, visualization; DM: methodology, writing, reviewing, and editing, supervision; AS: methodology, writing, reviewing, and editing, supervision.








\bibliographystyle{cas-model2-names}

\bibliography{cas}



\end{document}